\documentclass{article}
\usepackage{graphicx} 
\usepackage{mathrsfs}
\usepackage{mathrsfs}
\usepackage{amsfonts}
\usepackage{amsmath}
\usepackage{hyperref}
\usepackage{booktabs}
\usepackage[sort&compress,numbers]{natbib}
\usepackage{floatrow}
\usepackage{caption}
\usepackage{multirow,tabularx}
\usepackage{float}
\usepackage{algorithm}
\usepackage{algpseudocode}
\usepackage{lipsum}
\usepackage[english]{babel}
\usepackage{amsthm}
\usepackage{mathrsfs}
\usepackage{listings}
\usepackage{graphicx,xcolor}
\usepackage{amsmath,amssymb,amsfonts}
\hypersetup{hidelinks}
\usepackage{natbib}
\usepackage[margin=2.5cm]{geometry}
\usepackage{setspace}
\usepackage{tikz}
\DeclareMathOperator{\sgn}{sgn}
\allowdisplaybreaks[1]
\usepackage{amssymb}
\usepackage{float}
\usepackage{multirow}
\usepackage[all]{xy}
\usepackage{graphicx}
\usepackage{xcolor}
\usepackage{subcaption}
\usepackage[numbers]{natbib}

\newenvironment{keywords}{%
  \par\medskip\noindent
  \small
  \textbf{Keywords:}\ \ignorespaces
}{%
  \par\medskip
}
\newtheorem{theorem}{Theorem}
\newtheorem{lemma}{Lemma}
\newtheorem{definition}{Definition}
\newtheorem{remark}{Remark}

\title{Adaptive Penalized Likelihood method for Markov
Chains}
\author{Yining Zhou \\
Department of Computer Science, Mathematics, Physics and Statistics, \\ University of British Columbia, Kelowna, Canada V1V 1V7
\and Min Gao \\ 
School of Big Data and Statistics, 
Anhui University, Hefei, PR China 230601;\\ Department of Computer Science, Mathematics, Physics and Statistics, \\ University of British Columbia, Kelowna, Canada V1V 1V7
\and Yiting Chen \\ 
Department of Computer Science, Mathematics, Physics and Statistics, \\ University of British Columbia, Kelowna, Canada V1V 1V7
\and Xiaoping Shi \\
Department of Computer Science, Mathematics, Physics and Statistics, \\ University of British Columbia, Kelowna, Canada V1V 1V7}

\begin{document}
\maketitle
\begin{abstract}
Maximum Likelihood Estimation (MLE) and Likelihood Ratio Test (LRT) are widely used methods for estimating the transition probability matrix in Markov chains and identifying significant relationships between transitions, such as equality. However, the estimated transition probability matrix derived from MLE lacks accuracy compared to the real one, and LRT is inefficient in high-dimensional Markov chains. In this study, we extended the adaptive Lasso technique from linear models to Markov chains and proposed a novel model by applying penalized maximum likelihood estimation to optimize the estimation of the transition probability matrix. Meanwhile, we demonstrated that the new model enjoys oracle properties, which means the estimated transition probability matrix has the same performance as the real one when given. Simulations show that
our new method behave very well overall in comparison with various competitors. Real data
analysis  further convince the value of our proposed method.
\end{abstract}

\section{Introduction}
In stochastic process analysis, the Markov transition matrix plays a crucial role in describing the numerical significance of the transformational relationship between object states at discrete time steps \cite{karlin1975collection, Asmussen2003,  gagniuc2017markov}. Meanwhile, it is widely used across various disciplines, including medicine \cite{Bebu2018, zhang2013forecasting}, finance \cite{duan2001american, kharvi2022optimal}, and sports analysis \cite{hirotsu2002using}. To estimate transition probabilities hidden within observed data, statisticians commonly adopt advanced estimation techniques such as Maximum likelihood estimation (MLE) \cite{bartlett1951frequency, craig2002estimation}. Consider applications within the medical sector as a case in point, Bebu and Lachin  \cite{Bebu2018} used MLE to estimate transition probabilities between five retinopathy states to screen for retinopathy in type 1 diabetes (T1D). This facilitated the development of optimal screening schedules and significantly reduced monitoring costs.\\

Nevertheless, as a data-drive estimation methodology, the application of MLE on Markov chain has been criticized for its inherent limitations \cite{athreya1992bootstrapping, datta1992bootstrap}, namely, that based on the current observed data, it may result in a non-ergodic estimated Markov chain, despite the true model being ergodic. To address this issue, Kulperger \cite{kulperger1999countable} devised a new bootstrap algorithm to revise the MLE. Kulperger performed bootstrapping to establish an ergodic Markov transition matrix and treated it as tuning parameters, which were then combined into the original MLE to ensure that the estimated transition matrix is ergodic on its state space. On the other hand, Anderson and Goodman \cite{anderson1957statistical}, Billingsley and Patrick \cite{billingsley1961statistical} proved that the estimated transition matrix by MLE has the asymptotic normality. Furthermore, they proved that under the null hypothesis, such as the transition probability of a first-order chain is constant, the asymptotic distribution of the likelihood-ratio test (LRT) on Markov chain converges to the chi-square distribution, similar to Wilks’ theorem on LRT \cite{wilks1938large}. Subsequently, McQueen and Thorley \cite{mcqueen1991stock};  Jirasakuldech et al. \cite{jirasakuldech2023non} applied the LRT on Markov chain within the financial domain. For example, when considering the null hypothesis that partial transitions have equal probabilities, Jirasakuldech et al. \cite{jirasakuldech2023non} showed that the U.S. housing market displays a non-random walk behavior and indicated the possibility of a bubble. In other words, if the null hypothesis in LRT not rejected, the LRT also contributed to revise the original MLE estimated transition matrix. \\

The primary focus of this paper is to revise Maximum likelihood estimation on Markov chain. This problem arises from researching the transition probabilities between the four types of bases found in a DNA molecule. We employed MLE on observed co-occurring ACGT core sequences in plant promoters across four plants \cite{ACGTplantes} and constructed a estimated transition matrix with distinct transition probabilities. Subsequently, we set the null hypothesis that two extremely close estimated transition probabilities are equal, and the further LRT did not reject this hypothesis, thus modifying the MLE transition matrix. However, different from the two-state Markov chain from McQueen et al. \cite{mcqueen1991stock} and Jirasakuldech et al. \cite{jirasakuldech2023non}, the transition matrix for DNA is constructed as a four-state matrix. This implies that the choices for all possible equality hypotheses exponentially increase from the two-state model. Consequently, it becomes time-consuming to verify all hypotheses.  Therefore, the paper serves to propose a novel algorithm that accurately estimates the transition matrix. Moreover, if there are partially equality transitions in the real transition matrix, these transitions will also be identified that enjoy the same probability within the estimated transition matrix. In other words, this problem is akin to a variable selection problem.\\

In linear models, regularization methods are widely used to address the variable selection problem, such as lasso regression \cite{lasso}, ridge regression \cite{Hoerlridge} and Penalized likelihood estimation (PLM) \cite{eggermont2001maximum}. These methods introduce penalty term to improve parameter prediction from traditional ordinary least squares estimation (OLS), and lasso has been widely employed in various fields because of its sparsity in parameter estimation \cite{you2018modelling, jafar2023hypgb, ghosh2021efficient}. Notably, the consistency of variable selection with lasso has long been a subject of skepticism among statisticians. Fan and Li \cite{Jianqing2001} proposed that the traditional lasso might not be an oracle procedure, meaning that the estimated model may not perform as well as the real one. Subsequently, Zhao and Yu \cite{zhao2006model}, and Zou \cite{zou2006} proved that consistency of variable selection in lasso only works under certain conditions. To address this limitation, recent researchers have made significant progress in extending the lasso regression. For example, Zou and Hastie \cite{zou2005regularization} proposed the elastic net technique as a novel shrinkage method that extends from lasso. It's elastic net penalty ensures accurate prediction in sparse models and performs well in misclassification on microarray data. Wang et al. \cite{wang2007robust} devised another robust shrinkage method, which combines Least Absolute Deviations (LAD) in lasso and is defined as LAD-lasso. Furthermore, applying LAD-lasso on Chinese stock market data proved its accurate prediction capability. Notably, Zou \cite{zou2006} also proposed a new regularization method by introducing data-dependent weighted term on the penalty term to further refine the OLS. This revision avoids the equal penalization of parameters by the L1-penalty, and the modified lasso is referred to as adaptive lasso. Moreover, Zou also provided theoretical evidence that the adaptive lasso is an oracle procedure. Following this, researchers have further developed the adaptive lasso, such as in the Cox proportional hazards model by Zhang and Lu \cite{zhang2007adaptive}, and the adaptive group lasso by Wang and Leng \cite{wang2008note}. \\

Significantly, regularization methods developed based on linear models have been widely employed by statisticians in nonlinear models. Zhu and Liu  \cite{zhu2009estimating} applied adaptive lasso to Hidden Markov models (HMM) to enhance the accuracy of the estimated covariance matrix of Gaussian Markov random fields (GMRF). Kang and Song et al. \cite{kang2019bayesian} further expanded the adaptive group lasso from Wang et al. \citep{wang2008note} to Bayesian adaptive group lasso, being the first to introduce the Bayesian lasso technique to HMMs. Furthermore, Topi{\'c} et al. \cite{topic2021synthesis} established a neural network (NN) model and developed a synthetic driving cycles validation system based on Markov chain and a lasso-based approach. Zhou and Song \cite{zhou2023functional} applied soft-thresholding operator and adaptive group lasso that further developed hidden Markov model. According to our investigate, recent popular applications of regularization methods on Markov chain are highly correlated with Bayesian analysis and Markov Chain Monte Carlo (MCMC). Generally, there currently does not exist a regularization method that can be leveraged to directly estimate Markov transition probabilities and explore the underlying equality relationships between different transitions. Notable, recent revised lasso methods such as LAD-lasso \cite{wang2007robust}, lasso bootstrap estimates (Bolasso) \cite{bach2008bolasso}, and Weighted LAD-lasso \cite{arslan2012weighted} have all been proven to exhibit similar properties to adaptive lasso, enjoying the oracle properties. Consequently, we adopted adaptive lasso to revise the estimated Markov transition matrix obtained by MLE. \\

The paper proposes a novel methodology named Penalized Likelihood Estimation with Adaptive lasso on Markov chain (McALasso) to revises the Maximum Likelihood Estimation for Markov chain. This approach extends the regularization methods from linear model to Markov chain, simultaneously estimating the transition matrix and detecting equality relationships between distinct transitions. In Section \ref{section_basic_estimation}, we provide an overview of the performance of the MLE on Markov chain, discussing their contributions and limitations through simulated and real data, respectively. The regularization methods (lasso, adaptive lasso) extended to Markov chain will be presented in Section 3. We defined McALasso which based on adaptive lasso, and further demonstrated its oracle properties. Furthermore, we utilized the \texttt{optimize} module of the \texttt{scipy} package in Python to implement the aforementioned methods on Markov chain. We employed a simulation study to show the superior performance of McALasso over lasso and MLE in variable selection. Moreover, we applied these methods to the real dataset of co-occurring ACGT core sequences across four plants \cite{ACGTplantes}. Once again, the comparison results further emphasize McALasso's superiority in variable selection. The computational results are presented in Section \ref{section_results}. 

\section{Estimation of the parameter of Markov chain} \label{section_basic_estimation}

\subsection{Maximum Likelihood Estimation on Markov Chain} \label{section_mle}
A discrete homogeneous ergodic Markov chain $\{ X_{s},\ 0\leq s  \}$ with non-negative integer finite state space $M$ = $\{1, \cdots , m \}$, which inherent properties ensure the transition probabilities between each states always positive \cite{karlin1975collection}. This Markov chain leads to the construction of an $m \times m$ Markov transition matrix $P$ with transition probabilities $p_{ij}$, defined as:
\begin{equation}
    p_{ij} = P(X_1 = j\ |\ X_0 = i) > 0,\ \forall i,j \in M \label{ergodic_constraint}
\end{equation}

\begin{equation}
  P =  \begin{bmatrix}
            p_{11} & \dots & p_{1m} \\
            \vdots & \ddots & \vdots \\
            p_{m1} & \dots & p_{mm} \\
        \end{bmatrix}
\end{equation}\\

Suppose that a observed random multi-state Markov chain $S$ of length $N$ is given, the total occurrences of each transition within $S$ can be readily determined and represented as $n_{ij}$, where $\sum_{i,j=1}^m n_{ij} = N-1$. Additionally, let $\mathcal{N}$ denote the matrix of $n_{ij}$, which maintains the same dimensions as $P$.\\

On the other hand, it's a fundamental fact that for any $P$, the summation of each row in $P$ must equal 1. This property has been identified as the right stochastic matrix property by recent researchers \cite{SZABO2015320, gagniuc2017markov}. Mathematically, this can be expressed as:
\begin{equation}
    p_{i \cdot} = \sum_{j=1}^{m} p_{ij} = 1,\ \forall i \in M \label{TPM_constraint}
\end{equation}
with respect to this fact, by applying the Maximum likelihood estimation, the estimated transition matrix $\hat{P}$ will be computed as:
 \begin{align}
    &L(P) = \prod_{i,j=1}^{m} {p_i{}_j}^{n_i{}_j},\ \quad \forall i,j \in M \label{mle_likelihood}\\
    &\hat{P} = \arg\max_{P}\ L(P) \\
    &n_{i\cdot} = \sum_{j=1}^{m} n_i{}_j,\quad\quad\quad\quad \forall i \in M\\ 
    &\hat{{p}}_i{}_j  = \frac{n_i{}_j}{n_{i\cdot}}\label{MLE_closeform}
\end{align}

Notable, as a data-dependant methodology, the estimated transition probability $\hat{p}_{ij}$ is highly depend on the observation Markov sequence. In other words, if a particular transition is not observed, can lead to an incorrect estimation result for an ergodic Markov chain. Generally, let $P$ represents any Markov transition matrix. For $\forall\ i,j \in M$, the limitation of MLE can be expressed as:
\begin{align}
n_{ij} \neq 0\  &\rightarrow\ \ p_{ij} \neq 0  \\
n_{ij} = 0\  &\not\rightarrow\ \ p_{ij} = 0
\end{align}
where unobserved $n_{ij}$ occurs, Eq.~\ref{MLE_closeform} yields a result of 0, which contradicts the definition of ergodic Markov chain (Eq.~\ref{ergodic_constraint}). Moreover, in the scenario of an extreme observation sequence where $n_{i\cdot} = 0$, the Eq.~\ref{MLE_closeform} will produces an undefined outcome. To address this limitation, Kulperger \cite{kulperger1999countable} devised a bootstrapping algorithm that constructs an ergodic Markov transition matrix $Q$, which is treated as tuning parameter to revise the MLE as $\hat{P}_{(boost)}$.
\begin{remark} \label{remark_boostrap_mle}
Consider a Markov chain with a finite state set $M$. Let $Q = (q_{ij})$, where $Q$ is an ergodic Markov transition matrix. The formulation of $Q$ can be treated as a random walk on $M$. Let $\alpha \geq
 0$. Then $\hat{P}_{boost}$ is defined as:
        \begin{equation}
            \tilde{r}_{ij} = \begin{cases}
                                \frac{n_{ij}}{n_{i\cdot}} & \text{if } n_{i\cdot} \neq 0 \\
                                0 & \text{if } n_{i\cdot} = 0
                            \end{cases}
        \end{equation}
        \begin{equation}
            \hat{p}_{{ij}_{boost}} = \frac{\alpha q_{ij} + n_{ij}}{\alpha + n_{i\cdot}},\quad \hat{p}_{{ij}_{boost}}  \in \hat{P}_{boost}
        \end{equation}
\end{remark}

This lemma is quoted from Kulperger \cite{kulperger1999countable}. This revision ensures that MLE estimated Markov transition matrix always enjoys the property of ergodic Markov chain as the real one does. Meanwhile, if $\alpha = 0$, then $\hat{P}_{boost}$ is equal to $\tilde{R}=(\tilde{r}_{ij})$, which is equivalent to $\hat{P}$ by MLE.\\

Notably, Markov chain have been shown to possess a property similar to the multinomial distribution \cite{anderson1957statistical, whittle1955some}. This implies as $n$ increase,  it is difficult to identify underlying relationships between distinct transitions of the real model, such as equality. To verify this proposition, suppose that we are given a real 3-state $P^*$:

\begin{equation}
\centering
    {P^*} = \begin{bmatrix}
                0.4 & 0.1 & 0.5 \\
                0.45 & 0.3 & 0.25 \\
                0.2 & 0.1 & 0.7
            \end{bmatrix}
\end{equation}

where $p^*_{12} = p^*_{32} = 0.1$. Thus, we expect that $\hat{P}$ should also reflect this equality relationship. To validate this, we generated 1000 random sequences based on $P^*$ and computed each corresponding $\hat{P}$. The comparision results are presentes as follows:

\begin{table}[H]
    \centering
    \caption{Equality Detection on MLE }
    \begin{tabular}{c|ccccc}
    \toprule
    \textbf{Equality} & $\mathbf{N} $ & \textbf{Total} & \textbf{False} & \textbf{True}\\ \midrule
    $\hat{p}_{12} = \hat{p}_{32}$  & 10000 & 1000 & 1000 & 0\\
    \bottomrule
    \end{tabular}
    \label{Equality Detection on MLE}
\end{table}
\begin{figure}[H]
    \centering
        \includegraphics[width=8cm]{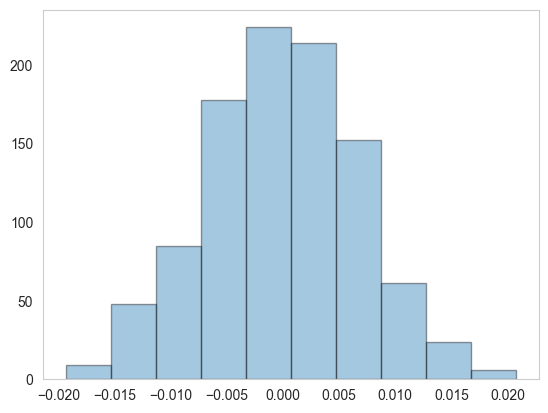}
        \caption{Distribution of Difference between $\hat{p}_{12} \And \hat{p}_{32}$}
        \label{Distribution of Difference between MLE}
\end{figure}

Based on Table.~\ref{Equality Detection on MLE}, the successful detection of equality by MLE is only 5\%, which is much lower than the failure detection rate (95\%). In conclusion, MLE underperforms in detecting underying equality relationships. Additionally, Figure.~\ref{Distribution of Difference between MLE} presents an interesting finding: the difference between $\hat{p}_{01}$ and  $\hat{p}_{21}$ nearly follows a Normal distribution (see Section.~\ref{section_lasso_adalasso} for more details and proof).\\

\subsection{Likelihood Ratio Test on Markov Chain} \label{section_lrt}
The true relationship between distinct transitions in any Markov chain cannot be directly observed. Therefore, conducting hypothesis testing based on the $\hat{P}$ provides an efficient approach. Anderson and Goodman \cite{anderson1957statistical} (also discussed in Billingsley \cite{billingsley1961statistical}) demonstrated that under certain null hypotheses, the likelihood-ratio test for a Markov chain is asymptotically distributed as a chi-square distribution. Furthermore, the equality assumption for distinct $p_{ij}$ has been employed in the financial field by McQueen et al. \cite{mcqueen1991stock}, Mills and Jordanov \cite{mills2003size}, and Jirasakuldech et al. \cite{jirasakuldech2023non}. In other words, the equality assumption is contributes to further revise $\hat{P}$ if the null hypothesis is true. \\

For verification purposes, we performed this method on the dataset pertaining to the ACGT repeat cis-element landscape across four plant genomes as provided by Mehrotra et al. \cite{ACGTplantes} (see Section.~\ref{section_results} for more details about the dataset). The ACGT data sequence is a classic Markov chain with 4 states corresponding to the bases of the DNA molecule. By randomly selecting a 10,000-length ACGT data sequence, then  $\mathcal{N}$ and $\hat{P}$ can be computed as follows:

$\mathcal{N}$ = \bordermatrix{ 
           & \text{A} & \text{C} & \text{G} & \text{T} \cr
            \text{A} & 896 & 478 & 625 & 927 \cr
            \text{C} & 665 & 462 & 218 & 579 \cr
            \text{G} & 645 & 440 & 466 & 531 \cr
            \text{T} & 720 & 543 & 774 & 1030 } 
\qquad
${\hat{P}}$ = \bordermatrix{ 
           & \text{A} & \text{C} & \text{G} & \text{T} \cr
           \text{A} & 0.306 & 0.163 & 0.214 & 0.317 \cr
           \text{C} & 0.346 & 0.240 & 0.113 & 0.301 \cr
           \text{G} & 0.310 & 0.211 & 0.224 & 0.255 \cr
           \text{T} & 0.235 & 0.177 & 0.252 & 0.336 }\\

In current $\hat{P}$, multiple pairs of $\hat{p}_{ij}$ values exhibit a high degree of similarity. Therefore, we established the null hypothesis based on the smallest difference, which assumes $p_{AG}$ and $p_{GC}$ to be equal. The hypothesis set is expressed as:
\begin{itemize}
    \item $H_0:$ ${p}_{AG}$ = ${p}_{GC}$
    \item $H_{a}: \forall\ i,j \in \{A,C,G,T\},\ p_{ij}$ are distinct 
\end{itemize}

Suppose that there is a set of classes $\mathcal{C}$ of $P$, which is defined as: $\mathcal{C} = \{\mathcal{S}_\rho,\ 0 <  \rho \leq 1 \}$, where $\mathcal{S}_\rho = \{(i,j) \ |\ p_{ij} = \rho \}$. Subsequently, the hypotheses $H_0$ and $H_{a}$ mentioned above will be reformulated as $\mathcal{C}_0$ and $\mathcal{C}_{\alpha}$, respectively, in the following manner:
\begin{itemize}
    \item $H_0:$ $\mathcal{C} = \mathcal{C}_0$
    \item $H_{a}:$ $\mathcal{C} = \mathcal{C}_{\alpha}$ 
\end{itemize}
Notably, the hard constraint (Eq.~\ref{TPM_constraint}) for any $P$ should always be satisfied, and the set of classes is equivalent to the parameter space. Meanwhile, $H_{a}$ assumes that all $p_{ij}$ are distinct, implying that $|\mathcal{C}_{\alpha}| = m^2$ with probability 1. Consequently, in hypothesis testing, it is always true that $|\mathcal{C}_{0}| < |\mathcal{C}_{\alpha}| = m^2$. Subsequently, the LRT for Markov chain is then defined as follows:
    \begin{equation}
        \Gamma = -2 \log \left[ \frac{\sup_{\mathcal{C} \in \mathcal{C}_{0}} L(\mathcal{C})}{\sup_{\mathcal{C} \in \mathcal{C}_{\alpha}} L(\mathcal{C})} \right]\ \rightarrow_d\ \chi^2_{(|\mathcal{C}_{\alpha}| - |\mathcal{C}_0|)}
    \end{equation} 
where the likelihood-ratio term in brackets is derived from two likelihood functions with their respective parameter spaces. The $\sup$ notation refers to computing the maximum value using the estimated parameters from MLE. Similarly to Wilks’ theorem \cite{wilks1938large}, by taking the difference between $\mathcal{C}_0$ and $\mathcal{C}_{\alpha}$ as the degrees of freedom of the $\chi^2$ distribution, we can determine whether to reject the null hypothesis. For any decision to fail to reject, the $\hat{P}$ will be revised by accepting that distinct $\{i,j\}$ enjoy the same transition probability.\\

 In consequence, by employing the commonly used significance level ($\alpha = 0.05$), the results of the LRT are presented in Table \ref{chitest_acgt}. The decision fail to reject the null hypothesis confirms the assumption of equality between $p_{AG}$ and $p_{GC}$ is true, which implies that $\hat{P}$ be revised to $\hat{P}_{LRT}$, as follows:
\begin{table}[H]
    \centering
    \caption{Likelihood ratio test in ACGT dataset}
    \begin{tabular}{c|cccc}
    \toprule
    \textbf{Null Hypothesis} & \textbf{df} & $\boldsymbol{\Gamma}$ & $\boldsymbol{\chi^2_{df}}$ & \textbf{Reject}\\ \midrule
    ${p}_{AG} = {p}_{GC}$    & 1 & 0.037 & 3.84 & \textbf{Fail}\\
    \bottomrule
    \end{tabular}
    \label{chitest_acgt}
\end{table}
${\hat{P}} = \bordermatrix{ 
           & \text{A} & \text{C} & \text{G} & \text{T} \cr
           \text{A} & 0.306 & 0.163 & \textbf{0.214} & 0.317 \cr
           \text{C} & 0.346 & 0.240 & 0.113 & 0.301 \cr
           \text{G} & 0.310 & \textbf{0.211} & 0.224 & 0.255 \cr
           \text{T} & 0.235 & 0.177 & 0.252 & 0.336 }$
\qquad
$\hat{P}_{LRT} = \bordermatrix{ 
           & \text{A} & \text{C} & \text{G} & \text{T} \cr
           \text{A} & 0.307 & 0.164 & \textbf{0.213} & 0.317 \cr
           \text{C} & 0.346 & 0.24 & 0.113 & 0.301 \cr
           \text{G} & 0.309 & \textbf{0.213} & 0.223 & 0.255 \cr
           \text{T} & 0.235 & 0.177 & 0.252 & 0.336 }$\\

Regarding the above example, various limitations of LRT become apparent, especially the inherent subjectivity problem in designing the null hypothesis. The purpose of this null hypothesis is to discern the underlying equality among distinct probabilities $p_{ij}$, thereby simplifying the set $\mathcal{C}$  to revise the $\hat{P}$. For instance, the assumption made in Table \ref{chitest_acgt} is predicated on the minimal difference between distinct $\hat{p}_{ij}$. However, when endeavoring to ascertain equality across other pairs of transitions, a fair criterion for determining whether each pair of $\hat{p}_{ij}$ can indeed be considered equal is lacking. In other words, a lenient criterion might lead to overestimation, and vice versa. Instead of relying solely on a criterion, testing each pair of transitions can also lead to precise detection. However, the general null hypothesis considers only one class, but it is not limited to cases where that class includes a maximum of two $p_{ij}$. Therefore, as the number of states $m$ increases, the total number of possible assumptions will increase exponentially. This also inherently leads to the inevitable escalation of the total execution time of testing to astronomically large values, resulting in unforeseen time costs.\\

Given the purpose indicated by the aforementioned null hypothesis, and the inherent limitations of LRT. Suppose that there exists a methodology both capable of accurately estimating \( P \) and comprehensively exploring all underlying equality between distinct transitions, it would provide more convincing numerical support for researchers in the field of Markov chain. In the next section, we propose a novel approach that achieves these purposes and further demonstrate its oracle properties.

\section{Regularization Methods} \label{section_lasso_adalasso}
\subsection{Objective} \label{section_obj}
We have shown that the LRT helps to uncover the latent equivalence $p_{ij}$ of Markov chain. However, if we treat the difference between each pair of distinct $p_{ij}$ as a variable, the general estimation problem is converted into a variable selection problem. Consequently, for any discrete homogeneous ergodic Markov chain we pursue a novel procedure ($\delta$) with the following objectives: 

\begin{itemize}
    \item The estimated Markov transition matrix $\hat{P}(\delta)$ should closely approximate the real Markov transition matrix $P^*$
    \item For any pair of distinct transitions $\{i, j\}$ that share the same $p^*_{ij}$, $\hat{p}_{ij}(\delta)$ should also reflect this relationship
\end{itemize}

\subsection{Lasso}
The variable selection problem has been widely researched in linear models, the regularization methods outperforms in addressing this problem, such as lasso from Tibshirani \cite{lasso}. Considering a linear model $\boldsymbol{y} = \mathbf{X}\boldsymbol{\beta^*} + \boldsymbol{\epsilon}$, where $\boldsymbol{\beta^*}  = \{\beta^*_0, \beta^*_1,..., \beta^*_k \}$ is the vector of the real parameter coefficients, $\boldsymbol{y}$ is the vector of the response variables, $\mathbf{X}$ is the predictor matrix with $k$ predictors and $n$ observations, and the matrix size is $n \times k$. The lasso is defined as:
\begin{equation}
\textbf{Lasso: }  \boldsymbol{\hat{\beta}}_L = \arg\min_{\boldsymbol{\beta}}  \|\boldsymbol{y} - \mathbf{X}\boldsymbol{\beta}\|^2 + \lambda\sum_{j=1}^{k}|\beta_j|  
\label{lasso}\\
\end{equation}
 where the second term of Eq.~\ref{lasso} is the penalty term, which is defined as ($L_1$). Moreover, $\lambda$ is a non-negative parameter that controls the shrinkage of the coefficients. Notably, the absolute value in $L_1$ induces sparsity under particular conditions  \cite{donoho2003optimally} . In other words, $L_1$ penalizes the coefficients of parameters that are less correlated with the response variable, driving them towards zero. This facilitates the removal of parameters from the full model if their coefficients become exactly zero. Furthermore, Fan and Li \cite{Jianqing2001} discussed the extension of the penalized least squares idea of lasso to likelihood-based models. 

 \begin{lemma} \label{pls}
     Let $p_{\lambda}(\cdot)$ denote the penalty function, which penalizes the coefficients with the same function by the threshold $\lambda$. Then, the penalized least squares (PLS) is defined as:
     \begin{equation}
         \boldsymbol{\hat{\beta}}_{PLS}  = \arg\min_{\boldsymbol{\beta}} \|\boldsymbol{y} - \mathbf{X}\boldsymbol{\beta}\|^2 + n\sum_{j=1}^{k}p_{\lambda}(|\beta_j|)
     \end{equation}
 \end{lemma}

 The Lemma \ref{pls} is quoted from Fan and Li\cite{Jianqing2001}. Notably, they mentioned that the same $p_{\lambda}(\cdot)$ for each parameter is not necessarily. Subsequently, they further consider the generalized linear model Scenario and devised the penalized likelihood estimation(PLE):

 \begin{lemma} \label{ple_jianqing}
 Let the density function of $x_i$ and $y_i$ is defined as $f_i(g(\mathbf{x_i}^T \boldsymbol{\beta}), y_i)$, which implies its likelihood function $\ell_i = \log(f_i)$. The PLE is expressed as:
    \begin{equation}
        \boldsymbol{\hat{\beta}}_{PLE} = \arg\max_{\boldsymbol{\beta}} \sum_{i=1}^{n} \ell_i(g(\mathbf{x_i}^T \boldsymbol{\beta}),\ y_i) - n\sum_{j=1}^{k}p_{\lambda}(|\beta_j|)
        \label{ple_lasso}
    \end{equation}
 \end{lemma}
where if we remove the second term, the Eq.~\ref{ple_lasso} is equivalent to the MLE. It inspired us to utilized our MLE (Eq.~\ref{mle_likelihood}) and lasso (Eq.~\ref{lasso}) to construct the penalized likelihood with lasso on Markov chain.\\

With respect to the hypothesis testing objective in Section~\ref{section_lrt}, the penalization should operate on the differences of $\hat{p}_{ij}$ for distinct $\{i,j\}$. In other words, the smaller difference between distinct $\hat{p}_{ij}$, the penalty term will exert greater force to shrink the difference in these pair of $p_{ij}$ towards 0. For any $m$-state Markov chain, let $\mathcal{T} = \{ (i_1, j_1,i_2,j_2)\ |\ \forall\ i_1 \neq i_2,\ j_1 \neq j_2,\ 1 \leq i_1, j_1,i_2,j_2 \leq m  \}$, where $|\mathcal{T}| = \mathbf{C}^{m^2}_2$.Furthermore, $\forall t \in \mathcal{T}$, let $\bar{p}_{t}$ indicate the gap of the transition probabilities between a pair of distinct transitions, which is denoted as $\bar{p}_{t} = p_{t_1,t_2} - p_{t_3,t_4}$ where $t = \{t_1,t_2,t_3,t_4\} \in \mathcal{T}$. The penalized likelihood estimation with lasso on Markov chain (McLasso) estimates the $\check{P} $ are given by:
\begin{align}
  \textbf{McLasso: } \check{P}  = \arg &\min_{{P}} \sum_{i,j=1}^{m} - n_{ij} \cdot \log(p_{ij}) + \lambda \sum^{\mathcal{T}}_{t} |\Bar{p}_t|
    \label{McLasso}\\
    \text{s.t:} &\left\{
        \begin{aligned}
         & 0 <\ {p}_{ij} \leq\ 1,\ {p}_{ij}\in {P} \\
         & \sum_{j=1}^m {p}_{ij} = 1,\ \forall i \in M 
        \end{aligned}
        \right.
\end{align} 
where $\lambda$ varies with $N$. Meanwhile, the objective function Eq.~\ref{McLasso} is a convex function, which means it will achieve the optimal value by employing the global minimizer. On the other hand, the penalty term may fulfills the objective of selecting $t$ for $|\bar{p}_{t}| = 0$. At this point, the general $p_{ij}$ estimation problem has been transformed into a nonlinear constrained optimization problem (see Section~\ref{section_results} for its numerical results).\\

As discussed in Section~\ref{section_lrt}, $P^*$ and the latent equality relationships between distinct transitions cannot be observed directly. Therefore, the estimation method serves as an oracle procedure. With respect to Fan and Li's \cite{Jianqing2001} pioneering contributions on the definition of oracle properties and Zou's \cite{zou2006} detailed summary of oracle procedures, we illustrate the oracle properties of Markov chains with the following example: suppose that McLasso (Eq.~\ref{McLasso}) is an oracle procedure, $\check{P}$ indicates that $\check{p}_{11} = \check{p}_{22}$, implying that McLasso predicts $p^*_{11} = p^*_{22}$ to hold, which is true in $P^*$. In conclusion, for any oracle procedure ($\delta$) for Markov chain, the following oracle properties must be satisfied:

\begin{definition} \label{Def_oracle}
Let $\mathcal{A}^* = \{t: \bar{p}^*_t \neq 0,\ t \in \mathcal{T} \}$, $\hat{\mathcal{A}} = \{t: \bar{\hat{p}}_t(\delta) \neq 0,\ t \in \mathcal{T} \}$ 
    \begin{itemize}
    \item $\textbf{Variable selection consistency: } \lim_{N \to \infty} P(\mathcal{A}^* = \hat{\mathcal{A}}) = 1$.
    \item \textbf{Asymptotic normality:} $ \sqrt{N}( {\hat{P}}_{\hat{\mathcal{A}}}  -  {P}^*_{\hat{\mathcal{A}}}) \rightarrow_d  \mathscr{N}(0,  {\Sigma}_{ {P}^*} )$, where $ {\Sigma}_{ {P}^*}$ is the covariance matrix based on the subset of $ {P}^*$.
    
\end{itemize}
\end{definition}

\subsection{Adaptive Lasso}
It is worth emphasizing that the lasso has been demonstrated to not always satisfy the oracle properties. Zhao and Yu \cite{zhao2006model} and Zou \cite{zou2006} demonstrated that the consistency of lasso's variable selection only hold under certain conditions. This concluusion further implies that the McLasso we constructed based on lasso may not fulfill the requirements of the oracle procedure.To address the inconsistency in variable selection of lasso, Zou \cite{zou2006} proposed the adaptive lasso and proved its satisfaction of the oracle properties:

\begin{lemma}
    Based on Eq.~\ref{lasso}, let $\gamma > 0$, $\boldsymbol{\widetilde{\beta}}$ be a $ \sqrt{n}$-consistent estimator for $\boldsymbol{\beta}^*$, and $\boldsymbol{\hat{w}} = 1/|\boldsymbol{\hat{\beta}}_{OLS}|^\gamma$, where:
    \begin{equation}
        \textbf{Aadaptive Lasso: }  \boldsymbol{\Tilde{\beta}}  =   \arg\min_{\boldsymbol{\beta}}  \|\boldsymbol{y} - \mathbf{X}\boldsymbol{\beta}\|^2 + \lambda\sum_{j=1}^{k} \hat{w}_j |\beta_j| 
    \end{equation}
\end{lemma}

The adaptive lasso provides us with the direction to modify McLasso, which is constructing data-dependent weighted terms to adaptively penalize each $|\bar{p}_t|$ in Eq.~\ref{McLasso}. The McLasso pursues the revision of $\hat{P}$ through variable selection, where the MLE plays the initial estimation role akin to OLS in Eq.~\ref{McALasso}. This proposition is similar to Tran et al. \cite{tran2012predictive}, where they established the weighted term with each predictor's MLE of coefficient ($\boldsymbol{\hat{\beta}}_{\text{MLE}}$). Furthermore, recalling the penalty term proposed in McLasso, the weighted term of adaptive lasso on Markov chains is defined as $\hat{w}_t = 1/|\bar{\hat{p}}_t|^{\gamma}$, where $\gamma > 0$ and $t \in \mathcal{T}$. In conclusion, McALasso is defined as:
\begin{align}
    \textbf{McALasso: } {\Tilde{P}}  = \arg &\min_{P} \sum_{i,j=1}^{m} - n_{ij} \cdot \log(p_{ij}) + \lambda \sum^{\mathcal{T}}_{t}\hat{w}_t |\bar{p}_t|     \label{McALasso}\\
    \text{s.t:} &\left\{
        \begin{aligned}
         & 0 <\ {p}_{ij} \leq\ 1,\ {p}_{ij}\in {P} \\
         & \sum_{j=1}^m {p}_{ij} = 1,\ \forall\ 1 \leq i \leq m 
        \end{aligned}
        \right.
\end{align} 
where $\tilde{\mathcal{A}} = \{t: \bar{\tilde{p}}_t \neq 0 \}$. The weighted term $\hat{w}_t$ in Eq.~\ref{McALasso} contributes to adaptive shrinkage of the difference between $\bar{p}_{t}$ based on $\bar{\hat{p}}_{t}$. For any $t \in \mathcal{T}$ where $\hat{w}_t$ is extremely large, it will lead to the penalty term returning a relatively tiny value regardless of $\bar{p}_t$ and $\lambda_{N}$. Conversely, if $\hat{w}_t$ is small, $\bar{p}_t$ will be penalized towards zero to achieve minimization.\\

As of now, the application of regularization methods is not limited to linear models. For example, Zhu and Liu \cite{zhu2009estimating}, Kang and Song et al. \cite{kang2019bayesian}, and Zhou and Song \cite{zhou2023functional} have successfully applied adaptive lasso in Hidden Markov Models. Moreover, Bayesian models and MCMC algorithms have also been widely combined with adaptive lasso \cite{kang2019bayesian, chin2016minimizing}. In general, We have not yet found an oracle procedure capable of identifying potentially equal transition probabilities in a Markov chain.  \\

On the other hand, as discussed by Zou \cite{zou2006}, the proper choice of $\lambda$ contributes to adaptive being an oracle procedure. Therefore, we adopt the assumptions of $\lambda$ proposed by Zou and further prove that McALasso also enjoys the oracle properties.0

\begin{theorem}
According to Definition.~\ref{Def_oracle}, the McALasso may enjoy the following oracle properties, where $\gamma > 0$, $\frac{\lambda}{ \sqrt{N}} \rightarrow 0 $ and $\lambda N^{\frac{\gamma-1}{2}} \rightarrow \infty$:
    \begin{itemize}
        \item \textbf{Asymptotic normality:} $ \sqrt{N}( {\tilde{P}}_{\mathcal{A}}  -  {P}_{\mathcal{A}}^*) \rightarrow_d \mathscr{N}(0, \boldsymbol{\Sigma})$.
        
        \item \textbf{Variable selection consistency:} $\lim_{N \rightarrow \infty} P(\tilde{\mathcal{A}} = \mathcal{A}) = 1$.
    \end{itemize}
\end{theorem}
where the definition of $\boldsymbol{\Sigma}$ and proofs are presented in Appendix \ref{append}.

\subsection{Selection of Optimal Regularization Parameter}
To utilize either McLasso (Eq. \ref{McLasso}) or McALasso (Eq. \ref{McALasso}), the optimal $\lambda$ needs to be selected initially. In this section, we employed the five-fold cross-validation technique to perform numerical analysis, as performed by Fan and Li \cite{Jianqing2001}. Notably, for the weighted term's parameter $\gamma$ in McLasso, we allowed cross-validation to identify the optimal $\gamma$ as well. However, conducting cross-validations simultaneously to determine both the optimal $\lambda$ and $\gamma$ is intricate and time-consuming. For simplification purposes, we set $\gamma$ equal to 1 as suggested by Zou \cite{zou2006}.\\

Suppose that we are given an $N$-length sequence $\{ X_{s},\ s \geq 0\}$ from an $m$-state Markov chain. We then split $X_{s}$ into $k$ sets with respect to the order of $s$. This allows us to generate the full dataset, denoted as $\mathbf{T} = \{T_1,\ldots,T_{k}\}$. For each performance evaluation of one fold during the five-fold cross-validation, we will pick one subset from $\mathbf{T}$ to be the test set ($T_{\kappa}$) without repetition, and the remaining sets will be training sets ($T_{\kappa}^C$), where $1 \leq \kappa \leq k$. Furthermore, we also given the set of $\lambda$ which is denoted as $\Lambda$, which means $\forall \lambda \in \Lambda$, we will perform the five-fold cross-validation (k=5) to get result CV($\lambda$). Generally, the optimal regularization parameter $\hat{\lambda}$ as:
\begin{equation}
\hat{\lambda} = 
\arg\min_{\Lambda} CV(\lambda)
= \arg\min_{\Lambda} \sum_{\kappa=1}^5 \sum_{n_{ij} \in T_{\kappa}} \sum_{\tilde{p}_{ij} \in T_{\kappa}^C} -n_{ij}^{(T_{\kappa})} \times \log \left(\tilde{p}_{ij}^{T_{\kappa}^C}(\lambda) \right)
\end{equation}
\section{Numerical Analysis} \label{section_results}
\subsection{Simulation Data Analysis} \label{Simulation Data Analysis}
As discussed in Section.\ref{section_lasso_adalasso}, the oracle procedure should discern the underlying equality relationship between the distinct$ \{i,j\}$.  Suppose that we have $P^*$ with four pairs of transitions characterized by identical probabilities. Subsequently, we utilized this matrix to generate a Markov sequence of length 50,000 and compute $\mathcal{N}$, where $P^*$ and $\mathcal{N}$ are defined as:

\begin{center}
   $\mathcal{N}$ =  \bordermatrix{ & 1 & 2 & 3 \cr
      1 & 8508 & 4277 & 8583  \cr
      2 & 6823 & 3985 & 2684 \cr
      3 & 6038 & 5230 & 3871} \qquad
$P^* $ = \bordermatrix{
        & 1 & 2 & 3 \cr
    1&  0.4 & 0.2 & 0.4 \cr
    2&  0.5 & 0.3 & 0.2 \cr
    3&  0.4 & 0.34 & 0.26
      } 
\end{center}

We utilized these matrices to perform McLasso and McALasso. The 5-fold Cross-Validation results are presented below:

\begin{figure}[H]
\centering
\begin{minipage}[t]{0.49\textwidth}
\centering
\includegraphics[width=8cm]{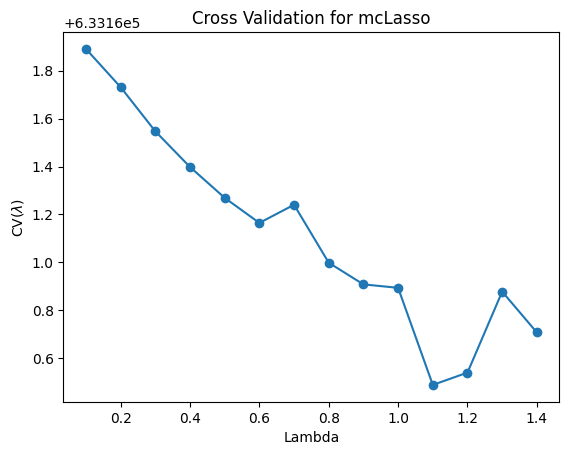}
\end{minipage}
\begin{minipage}[t]{0.49\textwidth}
\centering
\includegraphics[width=8cm]{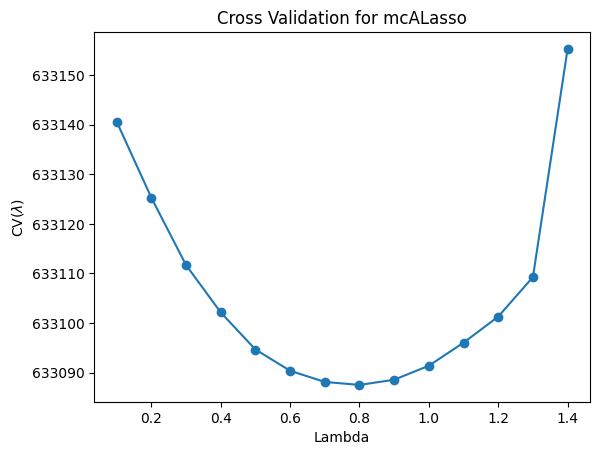}
\caption{Cross-Validation of McLasso and McALasso on Simulation data }
\label{cv_result}
\end{minipage}
\end{figure}
where we can observe that the two cross-validations successfully computed $\hat{\lambda}$ for both regularization methods for the same $\Lambda$, respectively. Subsequently, we employed $\hat{\lambda}_{\text{McLasso}}$ and $\hat{\lambda}_{\text{McALasso}}$ to compute $\check{P}$ and $\tilde{P}$ as follows:

\begin{center}
   $\check{P}$ =  \bordermatrix{
                & 1 & 2 & 3 \cr
      1 & 0.398 & 0.2   & 0.402  \cr
      2 & 0.505 & 0.295 & 0.199 \cr
      3 & 0.399 & 0.345 & 0.256} \qquad
    $\tilde{P}$ = \bordermatrix{
            & 1 & 2 & 3 \cr
        1&  0.4   & 0.201 & 0.4 \cr
        2&  0.504 & 0.295 & 0.201 \cr
        3&  0.4   & 0.345 & 0.255
          } 
\end{center}

Based on the above results, it is apparent that during the single test, the estimated transition matrix $ {\tilde{P}}$ is in closer proximity to the true transition matrix $ {P^*}$ when contrasted with the performance of $ {\hat{P}}$. Moreover, the capability to discern equality relationships between distinct transitions has been successfully demonstrated in $ {\tilde{P}}$, surpassing the performance of $ {\hat{P}}$.\\

To significantly demonstrate the superior performance of McALasso over transition matrix estimation and variable selection using MLE and McLasso, we further utilized the purity function \cite{manning2008introduction, shi2021exploring} and finite-dimensional complex norm space for comparison, respectively. As we discussed in Section.~\ref{section_lrt}, we can establish set of classes $\mathcal{C}$ for both $P^*$ and $\hat{P}(\delta)$, where $\delta$ indicates any estimation procedure. Let $\mathcal{C}^* = \{\mathcal{S}_{\rho^*},\ 0 <  \rho^* \leq 1 \}$, where $\mathcal{S}_\rho^*= \{(i,j) \ |\ p^*_{ij} = \rho^* \}$, and let $\hat{\mathcal{C}} = \{\mathcal{S}_{\hat{\rho}},\ 0 <  \hat{\rho} \leq 1 \}$, where $\mathcal{S}_{\hat{\rho}} = \{(i,j) \ |\ \hat{p}_{ij}(\delta) = \hat{\rho} \}$. To compute the purity values between $P^*$ and $\hat{P}(\delta)$, the purity measurement is defined as follows:
\begin{equation}
    \mathcal{S(\mathcal{C}^*, \hat{\mathcal{C}})} = \frac{1}{m^2} \sum_{j}^{|\hat{\mathcal{C}}|} \max_{1 \leq i \leq |\mathcal{C^*}|} |\mathcal{C}^*_i \cap \hat{\mathcal{C}}_j|
    \label{purity}
\end{equation}
where the purity value of 1 indicates perfect clustering, while a value of 0 suggests the opposite, indicating the worst clustering outcome.\\

On the other hand, purity does not contribute to detecting difference between $ {P^*}$ and $\hat{P}(\delta)$. Therefore, we utilized the finite-dimensional complex norm space to quantify the difference between $\hat{P}(\delta)$ and $ {P^*}$, where a lower norm value indicates smaller errors between these two transition matrices. Let $d(\cdot)$ indicates the norm function for the error between $\hat{P}(\delta)$ and $P^*$ for the same $\{ i,j \}$. The finite-dimensional complex norm space is defined as:
\begin{equation}
    d(P^*, \hat{P}(\delta)) = \sqrt{\sum_{i,j=1}^{m} [P^*_{i,j} - \hat{P}_{i,j}(\delta)]^2 }
    \label{norm}
\end{equation}

By employing 100 different seeds in Python, we generated 100 distinct sequences based on $P^*$ (Section.~\ref{Simulation Data Analysis}). Subsequently, we employed MLE, McLasso, and McALasso to estimate the transition matrices $  {\hat{P}}$, $  {\check{P}}$, and $  {\tilde{P}}$ for each sequence. Moreover, we utilized Eq.~\ref{purity} and Eq.~\ref{norm} to evluate the performance of each procedure. The results are presented as following tables:\\

\begin{table}[H]
\centering
\caption{Summary of Purity Values  for Different Procedures}
\label{purity_summary}
\begin{tabular}{@{}lllllll@{}}
\toprule
\textbf{Procedure} & \textbf{Min} & \textbf{1st Qu} & \textbf{Median} & \textbf{Mean} & \textbf{3rd Qu} & \textbf{Max} \\ \midrule
\textbf{MLE} & 0.2222 & 0.2222 & 0.2222 & 0.2222 & 0.2222 & 0.2222 \\
\textbf{McLasso} & 0.6667 & 0.6667 & 0.6667 & 0.7052 & 0.7778 & 0.8889 \\
\textbf{McALasso} & {0.6667} & \textbf{0.7778} & \textbf{0.8889} & \textbf{0.8669} & \textbf{1.0000} & \textbf{1.0000}
 \\ \bottomrule
\end{tabular}
\end{table}
 
\begin{table}[H]
\centering
\caption{Summary of Finite-dimensional Complex Norm Space for Different Procedures}
\label{norm_summary}
\begin{tabular}{@{}lllllll@{}}
\toprule
\textbf{Procedure} & \textbf{Min} & \textbf{1st Qu} & \textbf{Median} & \textbf{Mean} & \textbf{3rd Qu} & \textbf{Max} \\ \midrule
\textbf{MLE} & 0.0032 & 0.0083 & 0.0103 & 0.0104 & 0.0129 & 0.0173 \\
\textbf{McLasso} & 0.0032 & 0.0082 & 0.0103 & 0.0104 & 0.0128 & 0.0175 \\
\textbf{McALasso} & \textbf{0.0020} & \textbf{0.0067} & \textbf{0.0093} & \textbf{0.0091} & \textbf{0.0119} & \textbf{0.0169}
 \\ \bottomrule
\end{tabular}
\end{table}

Table \ref{purity_summary} indicates that MLE performs extremely poorly in detecting potential equality relationships between distinct transitions. McLasso shows some improvement over MLE but still falls short compared to McALasso. Additionally, Table \ref{norm_summary} reveals that the error between $  {\hat{P^*}}$ or $  {\check{P}}$ and $  {P^*}$ is similar, while $  {\tilde{P}}$ exhibits a smaller norm value compared to the other procedures. \\

\subsection{Real Data Analysis}
Now, let us consider the performance of the aforementioned regularization methods on a real dataset of co-occurring ACGT core sequences in plant promoters across four plants \cite{ACGTplantes}, which we utilized in Section~\ref{section_lrt}.\\

\begin{figure}[H]
\centerline{\includegraphics[scale=.6]{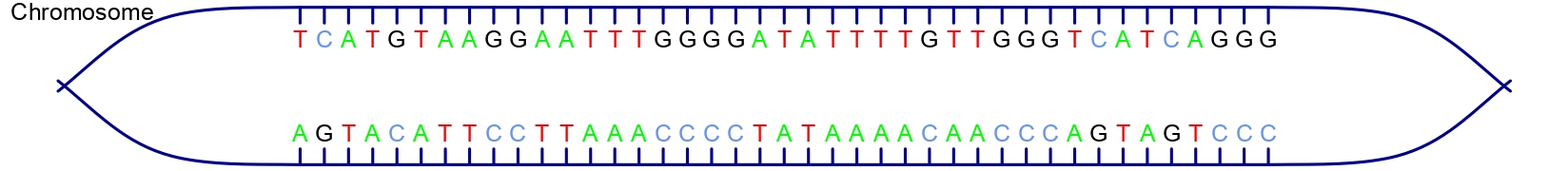}}
\caption{DNA Molecule Example}
\label{acgt_example}
\end{figure}

DNA is acknowledged as the hereditary material for all organisms. The DNA molecule consists of two intertwined strands, each linked together by bonds between bases. A sequence of bases in a portion of the DNA molecule, known as a gene, carries the instructions needed to assemble proteins. ACGT is an acronym for the four bases in the DNA molecule: Adenine (A), Cytosine (C), Guanine (G), and Thymine (T). One of the important discoveries is that A pairs with T and C pairs with G \cite{watson1953molecular}. Therefore, the study of any one single strand is sufficient to discover the pattern of transitions between the four bases. Figure~\ref{acgt_example} presents an example of a DNA molecule.\\

The real dataset that we tulized is the co-occurring ACGT core sequences in plant promoters across four plants: {Oryza sativa}, {Sorghum bicolor}, {Arabidopsis thaliana}, {Glycine max} \cite{ACGTplantes}. Given the substantial size of the original dataset, we employed a random selection process to extract a sequence length of 10,000 for the implementation of our devised regularization techniques. By intercepting the 500th to 600th positions of this sequence, the transfer tendencies among $\{A, C, G, T\}$ for the selected portion are presented as follows:

\begin{figure}[H]
\centerline{\includegraphics[scale=.8]{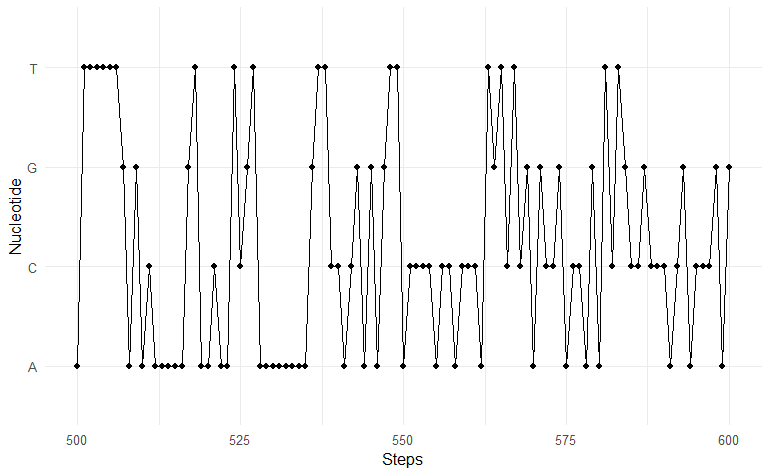}}
\caption{ACGT Molecule Example}
\label{DNA_example}
\end{figure}

On the other hand, the current transition $\{i,j\}$ is selected from $\{A, C, G, T\}$ which leads to a 4-state Markov chain. With respect to the selected sequence of length 10,000, for any fixed start state $i$, we can also observe the proportion of each next step's state $j$ over all possible next states. The results are presented as follows: 

\begin{figure}[H]
\centerline{\includegraphics[scale=.7]{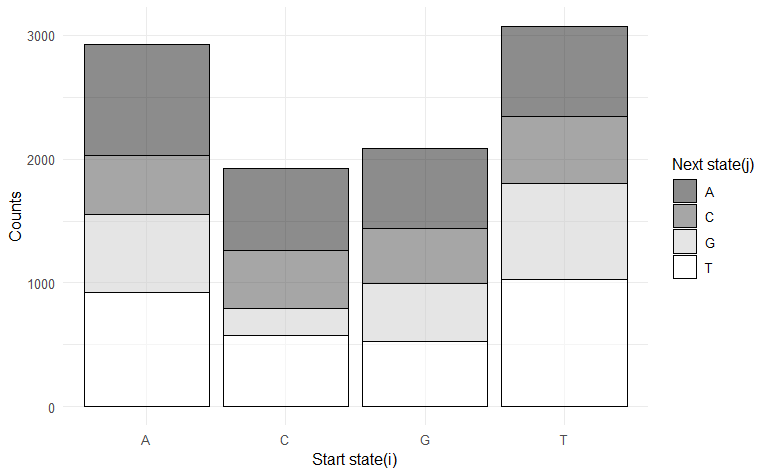}}
\caption{Transition Proportions for Each Nucleotide Pair in the ACGT Dataset
}
\label{DNA_dist_example}
\end{figure}

Furthermore, the numerical results in Figure~\ref{DNA_dist_example} validate our construction of $\mathcal{N}$. Meanwhile, $\mathcal{N}$ facilitates our utilization of MLE on the Markov Chain (Eq.~\ref{MLE_closeform}), enabling us to derive the initial estimation $\hat{P}$ as follows which are same as we did in Section.~\ref{section_lrt}:\\

$\mathcal{N}$ = \bordermatrix{ 
           & \text{A} & \text{C} & \text{G} & \text{T} \cr
            \text{A} & 896 & 478 & 625 & 927 \cr
            \text{C} & 665 & 462 & 218 & 579 \cr
            \text{G} & 645 & 440 & 466 & 531 \cr
            \text{T} & 720 & 543 & 774 & 1030 } 
\qquad
${\hat{P}}$ = \bordermatrix{ 
           & \text{A} & \text{C} & \text{G} & \text{T} \cr
           \text{A} & 0.306 & 0.163 & 0.214 & 0.317 \cr
           \text{C} & 0.346 & 0.240 & 0.113 & 0.301 \cr
           \text{G} & 0.310 & 0.211 & 0.224 & 0.255 \cr
           \text{T} & 0.235 & 0.177 & 0.252 & 0.336 }\\

Now, let us consider the performance of the aforementioned regularization methods based on $\mathcal{N}$ and $\hat{P}$. As discussed in Section~\ref{section_lrt}, the LRT demonstrates that some ACGT transitions might enjoy the same probability. However, the limitations of the LRT constrain its effectiveness in globally detecting all potential equality relationships. Therefore, we employed McLasso and McALasso to examine the same sequence that utilized in Section~\ref{section_lrt}. The five-fold cross-validation results are presented in following figure:

\begin{figure}[H]
\centering
\begin{minipage}[t]{0.49\textwidth}
\centering
\includegraphics[width=8cm]{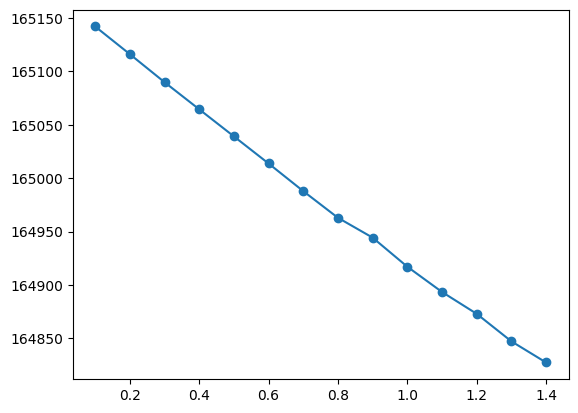}
\end{minipage}
\begin{minipage}[t]{0.49\textwidth}
\centering
\includegraphics[width=8cm]{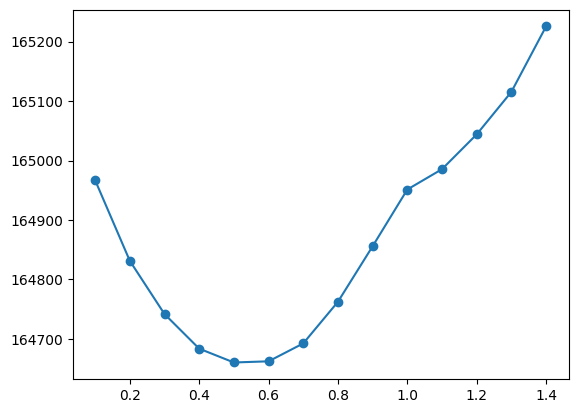}
\caption{Cross-Validation of McLasso and McALasso on ACGT Dataset }
\label{cv_real_result}
\end{minipage}
\end{figure}
where we can observe that for the same $\Lambda$, the cross-validations for McALasso can successfully compute its $\hat{\lambda}_{\text{McALasso}}$, whereas McLasso cannot. To detect the $\hat{\lambda}_{\text{McLasso}}$, we expanded its $\Lambda$. The results are presented as follows:

\begin{figure}[H]
\centerline{\includegraphics[scale=.6]{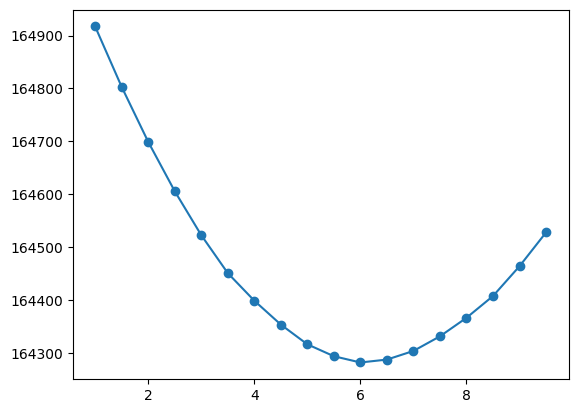}}
\caption{Cross-Validation of McLasso on ACGT Dataset with expanded $\Lambda$
}
\label{cv_real_expand_lasso_example}
\end{figure}

Subsequently, we employed $\hat{\lambda}_{\text{McLasso}}$ and $\hat{\lambda}_{\text{McALasso}}$ to compute $\check{P}$ and $\tilde{P}$ as follows:

${\check{P}}$ = \bordermatrix{ 
           & \text{A} & \text{C} & \text{G} & \text{T} \cr
           \text{A} & 0.303 & 0.169 & 0.218 & 0.311 \cr
           \text{C} & 0.335 & 0.244 & 0.121 & 0.300 \cr
           \text{G} & 0.303 & 0.217 & 0.227 & 0.253 \cr
           \text{T} & 0.237 & 0.182 & 0.253 & 0.329 }
\qquad
${\tilde{P}}$ = \bordermatrix{ 
           & \text{A} & \text{C} & \text{G} & \text{T} \cr
           \text{A} & 0.307 & 0.168 & 0.218 & 0.307 \cr
           \text{C} & 0.330 & 0.240 & 0.124 & 0.307 \cr
           \text{G} & 0.307 & 0.218 & 0.229 & 0.247 \cr
           \text{T} & 0.239 & 0.184 & 0.247 & 0.330 }\\

According to above results, $\check{P}$ shows that McLasso also has the capability to achieve the variable selection objective, such as $\check{p}_{AA} = \check{p}_{GA}$. Notably, $\tilde{P}$ also shows that equal transitions $\tilde{p}_{AA} = \tilde{p}_{GA}$ as McLasso and $\tilde{p}_{AG} = \tilde{p}_{GC}$ from LRT (Section~\ref{section_lrt}). Furthermore, McALasso can detect additional pairs transitions enjoy the same probability that $\check{P}$ is unable to identify, such as $\tilde{p}_{TG} = \tilde{p}_{GT}$ and $\tilde{p}_{AT} = \tilde{p}_{CT}$. Generally speaking, the real dataset further confirms the superior performance of McALasso in variable selection.

\section{Conclusion}
In this paper, we devised two penalized likelihood estimation methods that can be applied to Markov Chains: McLasso and McALasso with respect to Lasso and adaptive Lasso, respectively. Meanwhile, we showed that McLasso enjoys the oracle properties. By employing purity and the finite-dimensional complex norm space, we showed that McLasso is outperformed in parameter estimation and variable selection in data simulation. Moreover, the real data analysis further demonstrates the excellent performance of McALasso in variable selection.\\

The McLasso model implemented in Python is available on GitHub. The import sequences will be established with a finite number, where each distinct number indicates a distinct Markov Chain state. This model will adapt for any $m$-state Markov chain, where $m$ is a finite number. Therefore, this model can be flexibly applied across a wide range of areas, such as stock market prediction and disease progression screening. For instance, in the context of screening for clinical retinopathy progression, the transition probability matrix plays a crucial role in detecting the transition probability between each health state; see Bebu and Lachin’s paper \cite{Bebu2018}. Applying our model facilitates the further expansion of clinical retinopathy states and accurately determines whether transitions between different health states share the same transition probability. This approach enhances screening for disease progression and helps reduce screening budgets.

\break
\appendix
\section{Appendix} \label{append}
\textbf{Proof of Theorem 1}:
Based on our objective function of McALasso (Eq.~\ref{McALasso}), we first prove that this procedure enjoys the \textbf{asymptotic normality} property. Let $ {P} =  {P}^* + \frac{\mathbf{u}}{ \sqrt{N}}$, Eq.~\ref{McALasso} can be converted as:
\begin{equation*}
    \Psi (\mathbf{u}) =  \sum_{i,j=1}^{m} - n_{ij} \cdot \log(p^*_{ij} + \frac{u_{ij}}{ \sqrt{N}}) + \lambda \sum^{\mathcal{T}}_{t}\hat{w}_t |\Bar{p}^*_t + \frac{\bar{u}_t}{ \sqrt{N}}|
\end{equation*}

Let $\hat{\mathbf{u}}  = \arg \min_{u} \Psi $ which also implies that $ {\Tilde{P}}  =  {P^*} + \frac{\hat{\mathbf{u}} }{ \sqrt{N}}$ or $\hat{\mathbf{u}}  =  \sqrt{N}( {\Tilde{P}}  -  {P^*})$. Note that $\Psi (\mathbf{u}) - \Psi (\mathbf{0}) = \mathcal{V} (\mathbf{u})$, where $\mathcal{V} (\mathbf{u})$ is defined as:
\begin{equation*}
    \mathcal{V} (\mathbf{u}) \equiv    \sum_{i,j=1}^{m} \left\{ - n_{ij} \cdot \log(p^*_{ij} + \frac{u_{ij}}{ \sqrt{N}}) + n_{ij} \cdot \log(p^*_{ij}) \right\} +    
    \frac{\lambda}{ \sqrt{N}} \sum^{\mathcal{T}}_{t}\hat{w}_t  \sqrt{N} \left(|\Bar{p}^*_t + \frac{\bar{u}_t}{ \sqrt{N}}| - |\Bar{p}^*_t| \right)
\end{equation*}

We name the first summation term as $S_1$ and the sum of the penalty terms is named as $S_2$.  Then we do simplify $S1$ as:
\begin{align*}
    \centering
    S_1 &=  \sum_{i,j=1}^{m} - n_{ij}\cdot\log(p^*_{ij} + \frac{u_{ij}}{ \sqrt{N}})+n_{ij} \cdot \log(p^*_{ij}) \\
        &= \sum_{ij}^{m} - n_{ij}\cdot\log(\frac{p^*_{ij} + \frac{u_{ij}}{ \sqrt{N}}}{p^*_{ij}})\\
        &= \sum_{ij}^{m} - n_{ij}\cdot\log(1+\frac{u_{ij}}{{ \sqrt{N}p^*_{ij}}})\\
        &\approx  \sum_{ij}^{m} - n_{ij}\cdot \frac{u_{ij}}{{ \sqrt{N}p^*_{ij}}}
\end{align*}

If $\lambda = 0$, $\mathcal{V} (\mathbf{u})$ is equivalent to the MLE on Markov chain. Moreover, the Maximum likelihood estimation on Markov chain has been proved that it's enjoy the asymptotic normality property \cite{andreou2023bootstrap}. The asymptotic normality of MLE is defined as follows, which is quoted from AndReou \cite{andreou2023bootstrap}.
\begin{lemma}\label{asy_dist_mle}
Let the stationary distribution of $\{ X_{s},\  s \geq 0 \}$ denoted as $\pi$, where $\pi = \{\pi_1, \ldots, \pi_m\}$. Then $\hat{P} $ is asymptotically distributed as a normal distribution:
\begin{equation}
 \sqrt{N}( {\hat{P} } -  {P^*})\rightarrow_d \mathscr{N}(0,  \boldsymbol{\Sigma})
\end{equation}
\begin{equation}
   \boldsymbol{\Sigma}  = 
                \begin{bmatrix}
                    \frac{1}{\pi_1}\mathcal{Z}_1 & 0 & \cdots & 0 \\
                    0 &  \frac{1}{\pi_2}\mathcal{Z}_2 & 0 & 0 \\
                    \vdots & 0 &  \ddots & \vdots \\
                    0 & 0 & \cdots &  \frac{1}{\pi_m}\mathcal{Z}_m
                \end{bmatrix}_{m^2 \times m^2} 
\end{equation}
For any $i = {1,\ldots,M}$, then $\mathcal{Z}_i$ can be defined as:
\begin{equation}
   \mathcal{Z}_i  = 
        \begin{bmatrix}
            p_{i1}(1-p_{i1}) & -p_{i1}p_{i2} & \cdots & -p_{i1}p_{im} \\
            -p_{i1}p_{i2} & p_{i2}(1-p_{i2}) & \cdots & -p_{i2}p_{im} \\
            \vdots & \vdots & \ddots & \vdots \\
            -p_{i1}p_{im} &  -p_{i2}p_{im} & \ldots & p_{im}(1-p_{im})
        \end{bmatrix}_{m \times m}
\end{equation}
\end{lemma}

With respect the Lemma~\ref{asy_dist_mle}, it is sufficient to prove that $S_1$ is $O_p(1)$:
\begin{lemma}\label{lem5}
Let's consider ${\tilde{P}} = {P^*} + \frac{\mathbf{u}}{\sqrt{N}}$, where $\sum_{j=1}^m u_{i,j} =0$. Recall that $\hat{p}_{i,j}=n_{i,j}/n_{i\cdot}$. Then, we have
$$\sum_{i,j=1}^{m}  n_{ij}\cdot \frac{u_{i,j}}{{ \sqrt{N}p^*_{i,j}}}= \sum_{i,j=1}^{m} \hat{p}_{ij}n_{i\cdot}\cdot \frac{u_{i,j}}{{ \sqrt{N}p^*_{i,j}}}.$$
Proof. 
\begin{align*}
    \sum_{i,j=1}^{m}  n_{ij}\cdot \frac{u_{i,j}}{{ \sqrt{N}p^*_{i,j}}}&=
          \sum_{i,j=1}^{m}  \left(\hat{p}_{ij}-p^*_{i,j}+p^*_{i,j}\right)n_{i\cdot}\cdot \frac{u_{i,j}}{{ \sqrt{N}p^*_{i,j}}}\\
          & = \sum_{i,j=1}^{m}  \left(\hat{p}_{ij}-p^*_{i,j} \right)n_{i\cdot}\cdot \frac{u_{i,j}}{{ \sqrt{N}p^*_{i,j}}}+\sum_{i,j=1}^{m}   p^*_{i,j} n_{i\cdot}\cdot \frac{u_{i,j}}{{ \sqrt{N}p^*_{i,j}}} \\
        &= \sum_{i,j=1}^{m}  \left(\hat{p}_{ij}-p^*_{i,j} \right)n_{i\cdot}\cdot \frac{u_{i,j}}{{ \sqrt{N}p^*_{i,j}}}+\frac{1} {\sqrt{N}}\sum_{i }^{m}     n_{i\cdot}\cdot\sum_{j }^{m}  u_{i,j} \\
        &= \sum_{i,j=1}^{m}  \frac{\left(\hat{p}_{ij}-p^*_{i,j} \right)n_{i\cdot}}{ \sqrt{N}p^*_{i,j}}\cdot  u_{i,j},
\end{align*}
\end{lemma}
 where the coefficient $\left(\hat{p}_{ij}-p^*_{i,j} \right)n_{i\cdot}/(\sqrt{N}p^*_{i,j})$ is $O_p(1)$ because of $\hat{p}_{ij}-p^*_{i,j}=O_p(1/\sqrt{N})$ by Lemma \ref{asy_dist_mle}.\\

To interpret the convergence of $S_2$, recall $\mathcal{A}^* = \{t:  \bar{p}^*_t \neq 0 \}$. The convergence result depends on the following scenarios: $ t \in \mathcal{A}^*$ or  $t \notin \mathcal{A}^*$. Let's begin with the first scenario. If $t \in \mathcal{A}^*$, which implies $\bar{p}^*_t \neq 0$. Then we got the following revised $S_2$:

\begin{align*}
    S_2 &= \frac{\lambda}{ \sqrt{N}} \sum^{\mathcal{T}}_{t}\hat{w}_t  \sqrt{N} (|\Bar{p}^*_t + \frac{\bar{u}_t}{ \sqrt{N}}| - |\Bar{p}^*_t|)\\
        &= \frac{\lambda}{ \sqrt{N}} \sum^{\mathcal{T}}_{t}\hat{w}_t (|{\bar{u}_t} +  \sqrt{N}\cdot\Bar{p}^*_t| - | \sqrt{N}\cdot\Bar{p}^*_t|)
\end{align*}
Given $\frac{\lambda}{ \sqrt{N}} \rightarrow 0 $. Note that $ {\hat{w}_t} = 1/| {\bar{\hat{p}}_t}|^{\gamma} \rightarrow_p 1/| {\bar{p}^*_t}|^{\gamma}$ and $(|{\bar{u}_t} +  \sqrt{N}\cdot \bar{p}^*_t| - | \sqrt{N} \cdot \bar{p}^*_t|) \rightarrow 
 \bar{u}_t \cdot \sgn(\bar{p}^*_t)$. By Slutsky’s
theorem, $S_2 \rightarrow_p 0$ for $t \in \mathcal{A}^*$. In that scenario, $\mathcal{V} (\mathbf{u}) \equiv S_1 $, which implies $\mathcal{V} (\mathbf{u}) \rightarrow_d \mathscr{N}(0,  \boldsymbol{\Sigma})$.\\

In the second scenario, for $t \notin \mathcal{A}^*$, which implies $\bar{p}^*_t = 0$. Then, $S_2$ is defined as:
\begin{align*}
    S_2 &= \frac{\lambda}{ \sqrt{N}} \sum^{\mathcal{T}}_{t}\hat{w}_t (|{\bar{u}_t} +  \sqrt{N}\cdot\Bar{p}^*_t| - | \sqrt{N}\cdot\bar{p}^*_t|)\\
        &= \frac{\lambda}{ \sqrt{N}} \sum^{\mathcal{T}}_{t}\hat{w}_t \cdot |{\bar{u}}_t|\\
        &= \frac{\lambda}{ \sqrt{N}} \sum^{\mathcal{T}}_{t} \frac{1}{|{\bar{\hat{p}}_t}|^{\gamma}} \cdot |{\bar{u}}_t|\\
        &= \frac{\lambda}{ \sqrt{N}} N^{\frac{\gamma}{2}} \sum^{\mathcal{T}}_{t} \frac{1}{| \sqrt{N}\cdot{\bar{\hat{p}}_t}|^{\gamma}} \cdot |{\bar{u}}_t|
\end{align*}

Recall $\lambda N^{\frac{\gamma-1}{2}} \rightarrow \infty,\ \gamma >0$, and $| {\hat{P} } -  {P^*}| = O_p\left(\frac{1}{ \sqrt{N}}\right)$. Then for any $a \neq b$, let $\hat{p}_a$, $\hat{p}_b \in  {\hat{P} }$ and $p_a^*$, $p_b^* \in  {P^*}$, we have $|\hat{p}_a - p_a^* - (\hat{p}_b - p_b^*)| = O_p\left(\frac{1}{ \sqrt{N}}\right)$. Subsequently, suppose that $p_a^*, p_b^*$ follow the second scenario. Then we have $|\hat{p}_a - \hat{p}_b| = O_p\left(\frac{1}{ \sqrt{N}}\right)$, which means $\sqrt{N}|\hat{p}_a - \hat{p}_b| = O_p(1)$. Consequently, for any $t \in \mathcal{T}$, $1/| \sqrt{N}\cdot \bar{\hat{p}}_t\ |^{\gamma}$ will not converges to 0, where $\gamma > 0$. Moreover, $|{\bar{u}}_t|$ returns a constant value. By Slutsky’s theorem again,  the coefficient of $S_2 \rightarrow \infty$ for $t \notin \mathcal{A}^*$.\\

In conclusion, for every $\mathbf{u}$ we have $\mathcal{V} (\mathbf{u}) \rightarrow_d \mathcal{V}^*(\mathbf{u})$, where 

\begin{equation}
\mathcal{V}^*(\mathbf{u}) =  \left\{
    \begin{aligned}
     &\sum_{i,j=1}^{m} - n_{ij}\cdot \frac{u_{ij}}{\sqrt{N}p^*_{ij}}  & \quad t \in \mathcal{A}^* \\
     &\quad \infty\quad & \quad t \notin \mathcal{A}^* \\
    \end{aligned}
\right.
\label{asy_norm_p_tilde}
\end{equation}
where $\mathcal{V}^* (\mathbf{u})$ has the global minimizer $\mathbf{\hat{u}} = \arg \min_{\mathbf{u}} \mathcal{V} (\mathbf{u})$. By employing the epi-convergence results of Geyer \cite{geyer1994asymptotics, geyer1996asymptotics}, Knight and Fu \cite{fu2000asymptotics}, and Zou \cite{zou2006}, we have $\mathbf{\hat{u}}_{\mathcal{A}^*} \rightarrow_d \mathscr{N}(0, \boldsymbol{\Sigma})$ and $\mathbf{\hat{u}}_{\mathcal{A}^{*C}} \rightarrow_d 0$. At this point, we proved that McLasso enjoys the asymptotic normality property.\\

Now, let's prove that McLasso also enjoys variable selection consistent property. Consider the scenario that $t \in \mathcal{A}^*$ which implies $\bar{p}^*_t \neq 0$. Furthermore, Eq.~\ref{asy_norm_p_tilde} indicates $ \sqrt{N}( {\tilde{P}}_{\mathcal{A}^*}  -  {P}_{\mathcal{A}^*}^*) \rightarrow_d \mathscr{N}(0, \boldsymbol{\Sigma})$. So $|\bar{\tilde{p}}_t| \nrightarrow 0$ which means $P(t\in \tilde{\mathcal{A}}) \rightarrow 1$.\\

Conversely, for $\forall t^{\prime} \notin \mathcal{A}^*$, we need to show that $P(t^{\prime} \in \tilde{\mathcal{A}}) = 0$, where $\bar{p}^*_{t^{\prime}} = 0$ but $\bar{\tilde{p}}_{t^{\prime}} \neq 0$. By utilizing the same operational framework as Zou \cite{zou2006} via the Karush–Kuhn–Tucker (KKT) optimality condition to discuss the procedure's variable selection consistency property on  McALasso and let $\mathcal{F}$ indicates Eq.~\ref{McALasso}: 
\begin{equation}
    \frac{\partial \mathcal{F}}{\partial \bar{p}^*_{t^{\prime}}} \Rightarrow 
    \sum_{(i,j)\in t^{\prime}} -\frac{n_{ij}}{p^*_{ij}} \cdot \frac{\partial p^*_{ij}}{\partial \bar{p}^*_{t^{\prime}}} + \lambda \cdot \hat{w}_{t^{\prime}} \cdot \sgn(\bar{\Tilde{p}}_{t^{\prime}}) =0
\end{equation}
where the derivative is taken with respect to $\bar{p}^*_{t^{\prime}}$ for $\mathcal{F}$, where $t^{\prime} \in \tilde{\mathcal{A}}$. ${\partial \tilde{p}_{ij}}/{\partial \bar{p}^*_{t^{\prime}}}$ returns values of -1 or 1, similar to $\mathrm{sgn}(\bar{\tilde{p}}_{t^{\prime}})$.
Since $|\sum_{(i,j)\in t^{\prime}} -\frac{n_{ij}}{p^*_{ij}} \cdot \frac{\partial p^*_{ij}}{\partial \bar{p}^*_{t^{\prime}}}|\leq \sum_{i,j=1}^m  \frac{n_{ij}}{p^*_{ij}}=O_p(\sqrt{N})$ by Lemma \ref{lem5}. Note that $\frac{\lambda}{\sqrt{N}} \cdot \hat{w}_{t^{\prime}} = \frac{\lambda}{ \sqrt{N}} N^{\frac{\gamma}{2}} \cdot \frac{1}{| \sqrt{N}\cdot \bar{\hat{p}}_t|^{\gamma}} \rightarrow_p \infty$. In conclusion, for $\forall t^{\prime} \notin \mathcal{A}^*$, $P(t^{\prime} \in \tilde{\mathcal{A}})
\leq 
P\left(\sum_{(i,j)\in t^{\prime}} -\frac{n_{ij}}{p^*_{ij}} \cdot \frac{\partial p^*_{ij}}{\partial \bar{p}^*_{t^{\prime}}} + \lambda \cdot \hat{w}_{t^{\prime}} \cdot \sgn(\bar{\Tilde{p}}_{t^{\prime}}) =0\right)
\rightarrow
0
$.

\break

\end{document}